\DocumentMetadata{testphase={phase-III,math}}
\documentclass[12pt,letterpaper]{article}
\usepackage[utf8]{inputenc}
\usepackage[english]{babel}
\usepackage{amsmath}
\usepackage{amsfonts}
\usepackage{amssymb}
\usepackage{graphicx}
\usepackage{float}
\usepackage{listings}
\usepackage{color}
\usepackage[left=2cm,right=2cm,top=2cm,bottom=2cm]{geometry}
\usepackage[pdflang={en-US}, pdftitle={FDTD Model of Noise Pollution from an AI Data Center}]{hyperref}
\hypersetup{
    colorlinks,
    linkcolor={red!50!black},
    citecolor={blue!50!black},
    urlcolor={blue!80!black}
}
\usepackage{xcolor}
\usepackage{siunitx}
\DeclareSIUnit\year{yr}

\allowdisplaybreaks

\newcommand{\authorstyle}[1]{{\large\usefont{OT1}{phv}{b}{n}\color{black}#1}} 

\newcommand{\institution}[1]{{\footnotesize\usefont{OT1}{phv}{m}{sl}\color{black}#1}} 

\usepackage{titling} 

\newcommand{\HorRule}{\color{gray}\rule{\linewidth}{1pt}} 

\pretitle{
	\vspace{-30pt} 
	\HorRule\vspace{10pt} 
	\fontsize{32}{36}\usefont{OT1}{phv}{b}{n}\selectfont 
	\color{black} 
}

\posttitle{\par\vskip 15pt} 

\preauthor{} 

\postauthor{ 
	\vspace{-10pt} 
	\par\HorRule 
	\vspace{20pt} 
}

\title{FDTD Model of Noise Pollution\\ from an AI Data Center}

\author{
	\authorstyle{Gavin Buxton} 
	\newline\newline 
	\institution{Science Department, Robert Morris University, Moon Township, PA 15108, US.}\\ 
}

\date{}

\begin{document}

\maketitle 

\vspace{-50pt}
The sudden growth in AI data centers is raising numerous environmental concerns. Low-frequency noise pollution is a concern for local residents, that may not just be an annoyance but also carry significant health impacts.
An FDTD model of sound propagation over the topography near an AI data center is presented.
The topography plays a large role in the level of noise pollution, with areas of sudden elevation on either side of the river near the AI data center more susceptible to high noise levels, while areas further afield largely protected from noise pollution by the topography. 
Understanding the role that topography plays in noise pollution, and capturing the noise pollution in specific topographies, can help identify areas where noise mitigation strategies might be best employed.

\section*{Introduction}

Artificial Intelligence (AI) is becoming increasingly difficult to avoid in our everyday lives, while research on the environmental and human health impacts of AI data centers remains scarce \cite{george2026health}.
AI data centers are energy-intensive computing facilities critical to advanced AI applications, such as the training and inference of large language models \cite{george2026health, siddik2021environmental, kollar2025relationship}.
Already handling cloud computing, cryptocurrency, video streaming and other computational burdens \cite{siddik2021environmental}, data centers have grown exponentially as AI has transformed the way many live, work, and interact with technology \cite{Avelar2023ai, george2026health}.
Many important industries, from healthcare to manufacturing, are experiencing positive impacts from the use of pattern recognition and predictive algorithms behind the AI models \cite{Avelar2023ai}.
However, the rapid growth of AI has also caused an unprecedented increase in the electricity demand \cite{chen2025electricity}, which is expected to raise electricity costs  \cite{chen2025electricity}, increase emissions of fine particulate matter and carbon dioxide from power plants \cite{orgeval2024carbon, george2026health}, and stress the capacity of local grid infrastructure \cite{chen2025electricity}.
This energy is released as heat in the data centers, which to prevent hardware failures, require large scale cooling facilities.
In smaller systems, and in traditional data centers, air cooling was used \cite{Avelar2023ai}. But as the data centers grow, liquid cooling systems (which ultimately dissipates this heat through the evaporation of water from external cooling towers) are becoming increasingly prevalent \cite{Avelar2023ai, chen2025electricity}.
This requires large volumes of water, up to 5 million gallons per day per data center (equivalent to a town of up to 50,000 residents) \cite{george2026health}, primarily from municipal drinking water systems \cite{george2026health} and predominantly located in high water-stress regions already exacerbated by global warming \cite{siddik2021environmental, chen2025electricity, kollar2025relationship}.
Traditional data centers have been found to generate low frequency noise pollution exceeding 60 dB in the local community, that detrimentally impacts the health of local residents \cite{george2026health}.
This has been in large part due to the air conditioning equipment and cooling fans of air-cooled systems (which still represent 95\% of data centers) \cite{wheeler2026infrasound}, but the evaporative cooling towers associated with newer large-scale data centers will also cause significant noise pollution.
Sadly, while the potential financial benefits of AI are felt by distant shareholders, the location of AI data centers will see   
vulnerable lower-income communities disproportionately affected by the pollution, noise and water scarcity \cite{george2026health}.

Environmental noise can lead to non-auditory health effects, ranging from annoyance and sleep disruption to an elevated risk of cardiovascular disease \cite{basner2014auditory, razai2025impact}, which can effect residents in the vicinity of AI data centers \cite{wheeler2026infrasound}.
AI data centers may involve the continuous operation of cooling systems, generators, power transformers, heating systems, or adjacent energy generation facilities.
The persistent noise pollution from AI data centers is expected to be lower in frequency (or even infrasound), with reports as high as \qty{60}{dB} reported in nearby communities \cite{george2026health, wheeler2026infrasound}.
While it is not required for the noise to be consciously perceived to be detrimental \cite{wheeler2026infrasound}, such low frequency noise pollution is often perceived as a continuous hum \cite{george2026health}. 
Furthermore, episodic noise pollution has also been associated with AI data center operations \cite{george2026health} that will impact the potential annoyance and health impacts \cite{razai2025impact}.

Before turning our attention to potential health impacts of noise pollution it is worth noting that most research involves A-weighted noise monitoring and more audible-range noise sources (such as traffic) \cite{basner2014auditory, wheeler2026infrasound}.
A-weighting accounts for the fact that the human ear is less sensitive to lower frequencies, and attenuates low-frequency components to mimic how humans tend to perceive sound \cite{razai2025impact}.
However, a small proportion (approximately 2.5\%) of the population are believed to be more sensitive to these lower frequencies than others, with this sensitivity emerging later in life \cite{leventhall2004low}, which is perhaps why low-frequency noise comprise a significant proportion of reported complaints \cite{baliatsas2016health}.
While the use of A-weighted noise monitoring by AI data centers is a regulatory requirement, and not a deliberate prevarication, the predicted environmental and health consequences may be misleading.

The most common health outcome from AI data centers may be annoyance, mental health and a reduction in cognitive function.
Irritation, anger, and distress are the emotional states experienced by people annoyed by noise pollution \cite{razai2025impact}, with consistent evidence that low-frequency noise pollution causes such annoyance \cite{wheeler2026infrasound}.
Studies have also shown an association between mental health outcomes (anxiety, depression and medication usage) and noise pollution, but these are limited to traffic and aircraft noise \cite{razai2025impact, guha2022noise}. 
While the association between cognitive function (poorer reading ability in children or increased dementia risk in adults) and noise is mostly limited to intermittent noise pollution (such as aircraft noise near airports) \cite{razai2025impact, wheeler2026infrasound}, recently it was found that higher-order cognitive functions (such as reasoning and mathematical thinking) are particularly affected by either intermittent or continuous low-frequency noise \cite{liang2024effect}.
Intermittent and sudden noise pollution, which is not necessarily expected to be the case with the routine operation of AI data centers, have been associated with sleep disturbances, changes in heart rate and blood pressure, and increased risk of hypertension and ischaemic heart disease \cite{wheeler2026infrasound, razai2025impact, george2026health, munzel2021transportation}. 
While not specific to low-frequency noise pollution, the World Health Organization recommends keeping annual average outdoor nighttime noise levels below \qty{40}{dBA} to prevent adverse health effects, with indoor nighttime noise levels below \qty{30}{dBA} \cite{berglund2000new}. 
The guidelines, in A-weighted decibels, might underestimate the appropriate noise level when considering continuous noise pollution (as opposed to the intermittent noise pollution often used to derive such guidelines), but overestimate what is appropriate for low-frequency noises which are less attenuated by buildings \cite{wheeler2026infrasound} and not adequately characterized using A-weighted noise monitoring \cite{razai2025impact}.

Here, the finite-difference time-domain (FDTD) method is  used the capture the dynamic pressure-velocity equations of sound waves propagating from a central region over realistic terrain.
The site of an AI data center currently under construction in Beaver County, PA and the surrounding terrain constitutes the computational domain and the effects of acoustic refraction and vegetation are considered. 
The predicted noise level within \qty{2.5}{\km} of the AI data center are calculated.
The following section details the numerical approach used in this study.

\section*{Model}

The elevation around the AI data center location was obtained from cycle 1 LiDAR Digital Elevation Model (DEM) data for the southern zone of Pennsylvania \cite{pamap_dem_2008}. Figure 1a shows the elevation in a $\qty{5}{\km}\times\qty{5}{\km}$ square region surround the AI data center, and the angles used to obtain one-dimension interpolations of elevation as a function of distance from the AI data center.
Given the spatial resolution required, even for low frequency sound waves, a full three-dimensional FDTD simulation of the entire area would be computationally prohibitive. Therefore, the propagation of sound in two-dimensional cross-sections through the system is simulated using the FDTD at different angles. 
These angles are also depicted in Figure 1a.
The system size in two-dimensions is 29154$\times$3000, with a spatial resolution of $\Delta x = \qty{0.1715}{\m}$ and a time step of $\Delta t = \qty{1e-4}{\s}$. This ensures numerically stability and that the wavelength of the sound waves corresponds to at least 20 spatial steps.

The FDTD equation for updating the local pressure is of the form \cite{van2003numerical, de2008finite, lacanna2014influence}
\begin{equation*}
p_{i,j}^{n+1} = p_{i,j}^n - \acute{\rho}_{i,j} \acute{c}_{i,j}^2 \Delta t \left[\frac{v_{x, i+1,j}^{n+\frac{1}{2}} - v_{x, i,j}^{n+\frac{1}{2}}}{\Delta x} + \frac{v_{y, i,j+1}^{n+\frac{1}{2}} - v_{y, i,j}^{n+\frac{1}{2}}}{\Delta y} \right]
\end{equation*}
where $p_{i,j}^n$ is the pressure at the location $i,j$ and at the time step $n$, $\acute{\rho}_{i,j}$ is the local modified density, $\acute{c}_{i,j}$ is the local modified speed of sound, $\Delta t$ is the time step, $v_{x, i,j}^{n+\frac{1}{2}}$ is the velocity in the x-direction at the location $i,j$ and at the time step $n+\frac{1}{2}$, and $\Delta x$ is the spatial resolution.

The velocity in the x- and y-directions is updated using the following equations \cite{van2003numerical, de2008finite, lacanna2014influence}
\begin{equation*}
v_{x,i,j}^{n+\frac{1}{2}} = v_{x,i,j}^{n-\frac{1}{2}} - \frac{\Delta t}{\acute{\rho}_{i-\frac{1}{2}, j}} \left[ \frac{p^n_{i,j} - p^n_{i-1,j}}{\Delta x}  - R v_{x,i,j}^{n-\frac{1}{2}}\right]
\end{equation*}

\begin{equation*}
v_{y,i,j}^{n+\frac{1}{2}} = v_{y,i,j}^{n-\frac{1}{2}} - \frac{\Delta t}{\acute{\rho}_{i, j-\frac{1}{2}}} \left[ \frac{p^n_{i,j} - p^n_{i,j-1}}{\Delta y} - R v_{y,i,j}^{n-\frac{1}{2}}\right] 
\end{equation*}
where a damping term is included to capture the propagation of sound waves through vegetation; the flow resistivity, $R$, has units of \unit{\kilo\Pa\s\per\square\m}.

In order to ensure stability in the system the following condition should be enforced:
\begin{equation*}
\Delta t < \frac{\Delta x}{\sqrt{2} c}
\end{equation*}
The density of the air, $\rho$, depends on the temperature, $T$, and is given by
\begin{equation*} 
\rho = \frac{P}{R_{air}T}
\end{equation*}
where $P = \qty{101.325}{\kilo\Pa}$ is the atmospheric pressure and $R_{air} = \qty{287.053}{\J\per\kg\per\K}$ is the gas constant for air.
In addition, the speed of sound also depends on the air temperature, and is given by
\begin{equation*} 
c = 331.32 \sqrt{1 + \frac{T_C}{273.15}}
\end{equation*}
where $T_C = T - 273.15$ is the temperature in Celsius.
This allows us to capture acoustic refraction as a consequence of air temperature gradients in the atmosphere.

To capture the propagation of sound waves through porous media, such as forested areas surrounding the AI data center, the density and speed of sound can be modified and a damping term can be included \cite{van2014efficient}.
The modified density is given by
\begin{equation*}
\acute{\rho} = \frac{\rho \tau}{\phi}
\end{equation*}
where $\rho$ is the density of air, $\tau$ is the tortuosity, and $\phi$ is the porosity of the media that the sound wave is propagating through. Similarly the speed of sound is reduced, as $\tau$ is increased:
\begin{equation*}
\acute{c} = \frac{c}{\sqrt{\tau}}
\end{equation*}
If the value of porosity is known, then tortuosity can be estimated using empirical formula \cite{attenborough1993models}.
 \begin{equation*} 
\tau = \phi^{-0.5}
\end{equation*}
where, in the current model, we assume the porosity varies linearly from 0.5 at the ground to 1 at a height of \qty{15}{\m}. This is only in regions covered in vegetation as estimated from satellite imagery. While the average height and porosity will vary for different tree species \cite{hagen1971windbreak} and the season, these are considered to be typical values.
We assume the flow resistivity also decreases linearly with height in areas with trees, and will investigate the effects of this term in the next section.

In order to terminate the FDTD model we require absorbing layers at the boundary that restrict reflection. 
The perfectly matched layers (PML) boundary conditions are the standard manner of achieving this \cite{berenger1994perfectly, van2003numerical}.
Near the boundary, 300 lattice steps deep, is a region where the waves propagating in from the simulation domain are gradually absorbed with little reflection to mimic an infinitely large domain.
The acoustic pressure is artificially split into components normal and parallel to the boundary, which ensures plane waves are fully transmitted at all angles of incidence while the sound waves are absorbed as they propagate through this media.
The update equations for the pressure are given by
\begin{equation*}
p_{x, i,j}^{n+1} = p_{x, i,j}^n - \acute{\rho}_{i,j} c_{i,j}^2 \Delta t \left[\frac{v_{x, i+1,j}^{n+\frac{1}{2}} - v_{x, i,j}^{n+\frac{1}{2}}}{\Delta x}\right] - \Delta t k_{1, x} p_{x, i,j}^n
\end{equation*}
and
\begin{equation*}
p_{y, i,j}^{n+1} = p_{y, i,j}^n - \acute{\rho}_{i,j} c_{i,j}^2 \Delta t \left[\frac{v_{y, i,j+1}^{n+\frac{1}{2}} - v_{y, i,j}^{n+\frac{1}{2}}}{\Delta y}\right] - \Delta t k_{1, y} p_{y, i,j}^n
\end{equation*}
where $k_{1, x}$ and $k_{1, y}$ are damping coefficients, depending on which direction is parallel and which is perpendicular to the boundary.
The two pressures sum to regular pressure.
\begin{equation*}
p_{i,j}^n = p_{x, i,j}^n + p_{y, i,j}^n
\end{equation*}
The velocity equations in the PML region are given by
\begin{equation*} 
v_{x,i,j}^{n+\frac{1}{2}} = v_{x,i,j}^{n-\frac{1}{2}} - \frac{\Delta t}{\acute{\rho}_{i-\frac{1}{2}, j}} \left[ \frac{p^n_{i,j} - p^n_{i-1,j}}{\Delta x} \right] - \frac{\Delta t}{\acute{\rho}_{i-\frac{1}{2}, j}} k_{2, x} v_{x,i,j}^{n-\frac{1}{2}}
\end{equation*}
\begin{equation*} 
v_{y,i,j}^{n+\frac{1}{2}} = v_{y,i,j}^{n-\frac{1}{2}} - \frac{\Delta t}{\acute{\rho}_{i, j-\frac{1}{2}}} \left[ \frac{p^n_{i,j} - p^n_{i,j-1}}{\Delta y} \right] - \frac{\Delta t}{\acute{\rho}_{i-\frac{1}{2}, j}} k_{2, y} v_{y,i,j}^{n-\frac{1}{2}}
\end{equation*}
where $k_{2, x}$ and $k_{2, y}$ are damping coefficients. The values of these damping constants in the direction parallel to the PML are zero, $k_{2,\parallel} = k_{1, \parallel} = 0$, and in the perpendicular direction we take $k_{1, \perp} = k_{2,\perp} / \rho$. $k_{1, \perp}$ and $k_{2,\perp}$ increase with distance from the interface, to ensure the wave is increasingly absorbed while minimizing reflections.
In particular,
\begin{equation*} 
k_{1, \perp}(x) = k_{1, \perp, \text{max}} \left(\frac{x}{L} \right)^m
\end{equation*}
where $x$ is the location from the interface, $L$ is the total thickness of the PML and $m$ is a number between 3 and 4 (currently taken to be 3.5).

The source of the sound wave is taken to be a region \qty{30}{\m} high on top of a \qty{5}{\m} by \qty{5}{\m} structure, to mimic the emission from the rooftop of an AI data center.
The source is a \qty{100}{\Hz} sound wave in a Gaussian envelope given by
\begin{equation*} 
v_{i,j}^{n+\frac{1}{2}} = A_0 \sin \left( 2\pi f \cdot t\right) \exp\left(-\frac{(t - t_0)^2}{2 \sigma_t^2}\right)
\end{equation*}
where $A_0$ is the amplitude, $f$ is the frequency, $t_0 = \qty{0.5}{\s}$ is the time at the center of the pulse, and $\sigma = \qty{0.2}{\s}$ the standard deviation of the Gaussian envelope.
The average sound intensity that an area is exposed to can be obtained from
\begin{equation*} 
I = \frac{p_{rms}^2}{\rho c}
\end{equation*}
where $p_{rms}$ is root-mean-square pressure given by
\begin{equation*} 
p_{rms} = \sqrt{\frac{1}{t_T} \int_0^{t_T} p^2(t)\,dt}
\end{equation*}
where $t_T$ is the total simulated time and the integration is over time.
The intensity in the current simulations is from a two-dimensional wave which geometrically spreads as the inverse to distance, and not the inverse to distance squared. To mimic spherical spreading, the intensity is modified 
\begin{equation*} 
\acute{I}(r) = I(r) \dfrac{1}{a |r|}
\end{equation*}
where $\acute{I}(r)$ is the modified intensity that mimics spherical spreading, $I(r)$ is the calculated intensity a distance of $r$ from the AI data center, and $a$ is a constant that can be used to vary the level of sound. In the current model we assume the level of sound \qty{10}{\m} from the source is \qty{100}{dB}.

The above model is now used to predict the noise pollution in the vicinity of the AI data center, as a function of the local topography, the damping of acoustic waves by vegetation and atmospheric temperature gradients.

\begin{figure}[H]
\begin{center}
	\includegraphics[width=0.55\linewidth, alt = {A graph that shows the sound level and elevation as a function of distance from the AI data center.}]{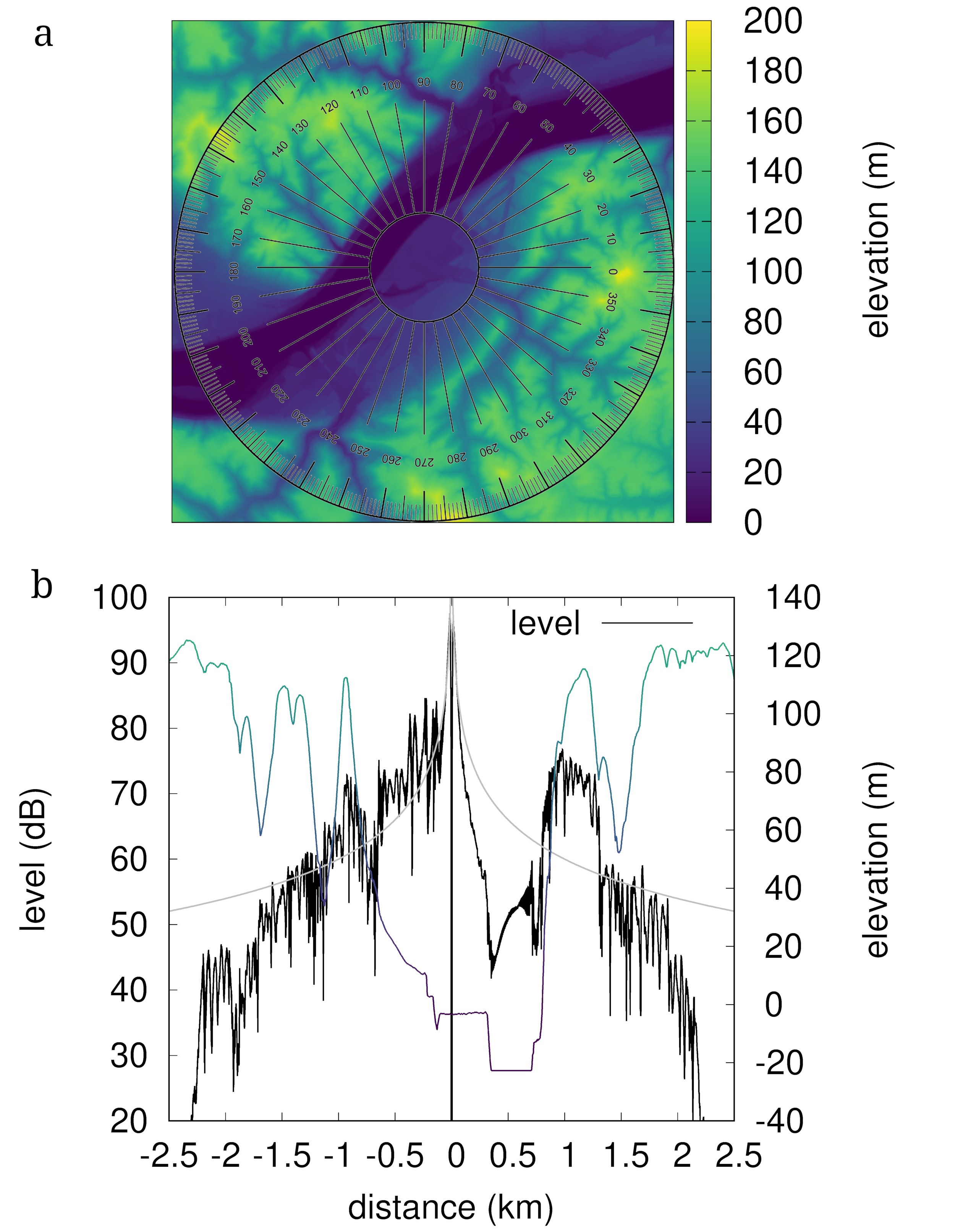} 
	\caption{a) The elevation in a $\qty{5}{\km}\times\qty{5}{\km}$ region surrounding the AI data center and a visual representation of the angles of one-dimensional profiles used in the two-dimensional FDTD model.  b) The predicted sound level and elevation as a function of distance from the AI data center. The sound is assumed to be \qty{100}{dB} a distance of \qty{10}{\m} from the center of the site, and the effects of vegetation and temperature gradients are not included. The elevation data (and coloring) correspond with the above contour plot at an angle of \qty{140}{\degree} (extending through the center such that negative distances are towards \qty{320}{\degree}).}
\end{center}
\end{figure}

\section*{Results}

The predicted sound level at an angle of \qty{140}{\degree} from the AI data center is plotted in Fig. 1b. The black line is the sound level in decibels, the light gray line represents the spherical spreading of sound waves, and the colored line is the elevation, as a function of distance from the AI data center. The AI data center is near the Ohio river, which is the low region of elevation at a distance of \qty{0.5}{\km}. 
On the other side of the river is Industry, PA (positive distance) and in the opposite direction is Raccoon Township, PA (negative distances). Note the elevation increases sharply either side of the river and the site of the AI data center.
This is reflected in the level of noise pollution.
The sound waves traveling from the AI data center, and reflecting from other surfaces, impact these regions of sudden elevation and result in localized regions of higher noise pollution. 
Higher than would be expected from simple spherical spreading alone.
However, the other side of these sudden increases in elevation, where regions close to the ground are not in a direct line of sight to the AI data center, the noise level drops off considerably.

This can be further seen in Fig. 2 which depicts a contour plot of the sound level in a circular region of radius \qty{2.5}{\km} around the AI data center. 
The two-dimensional contour data is interpolated from 180 FDTD simulations at each angle (from \qty{0}{\degree} to \qty{179}{\degree}) where negative distances in those simulation cover the angles from \qty{180}{\degree} to \qty{359}{\degree}.
Therefore, reflections out of the plane of the simulations (in directions perpendicular to the radial and vertical directions) are not captured. It is assumed such effects will be small.
Regions of elevation experience higher noise pollution, while dips in the topography can experience very low noise pollution.
Interestingly the level of noise pollution down the river and towards Midland, PA or Beaver, PA is not predicted to be as significant as one might fear, given the reduced elevation in these directions.

\begin{figure}
\begin{center}
	\includegraphics[width=\linewidth, alt = {A 3D contour plot showing the sound level as color on top of a topographical surface.}]{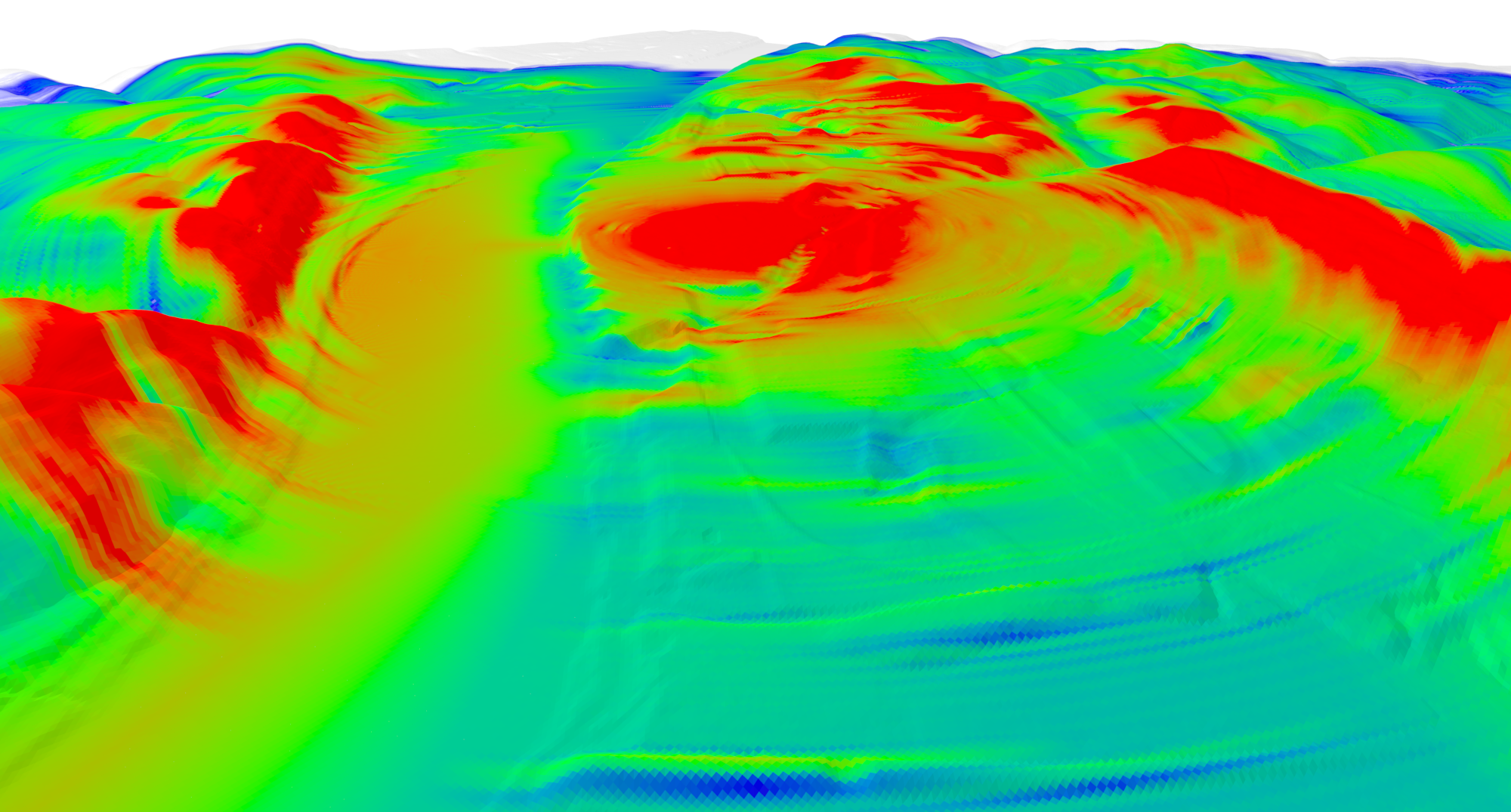} 
	\caption{The predicted sound level as a function of distance from the AI data center. The two-dimensional contour is interpolated from 180 FDTD simulations; a 2D simulation at every angle from the center of the site. The sound is assumed to be \qty{100}{dB} a distance of \qty{10}{\m} from the center of the site, and the effects of vegetation and temperature gradients are not included. Red is \qty{80}{dB}, green is \qty{53.3}{dB} and blue is \qty{26.6}{dB}.}
\end{center}
\end{figure}

\begin{figure}[h]
\begin{center}
	\includegraphics[width=0.55\linewidth, alt = {Two plots. The top plot shows how the porosity and modified density change as a function of height to mimic trees thinning near their tops, and the second plot is the sound level as a function of distance from the site as a function of different flow resistivity values at the base of the trees.}]{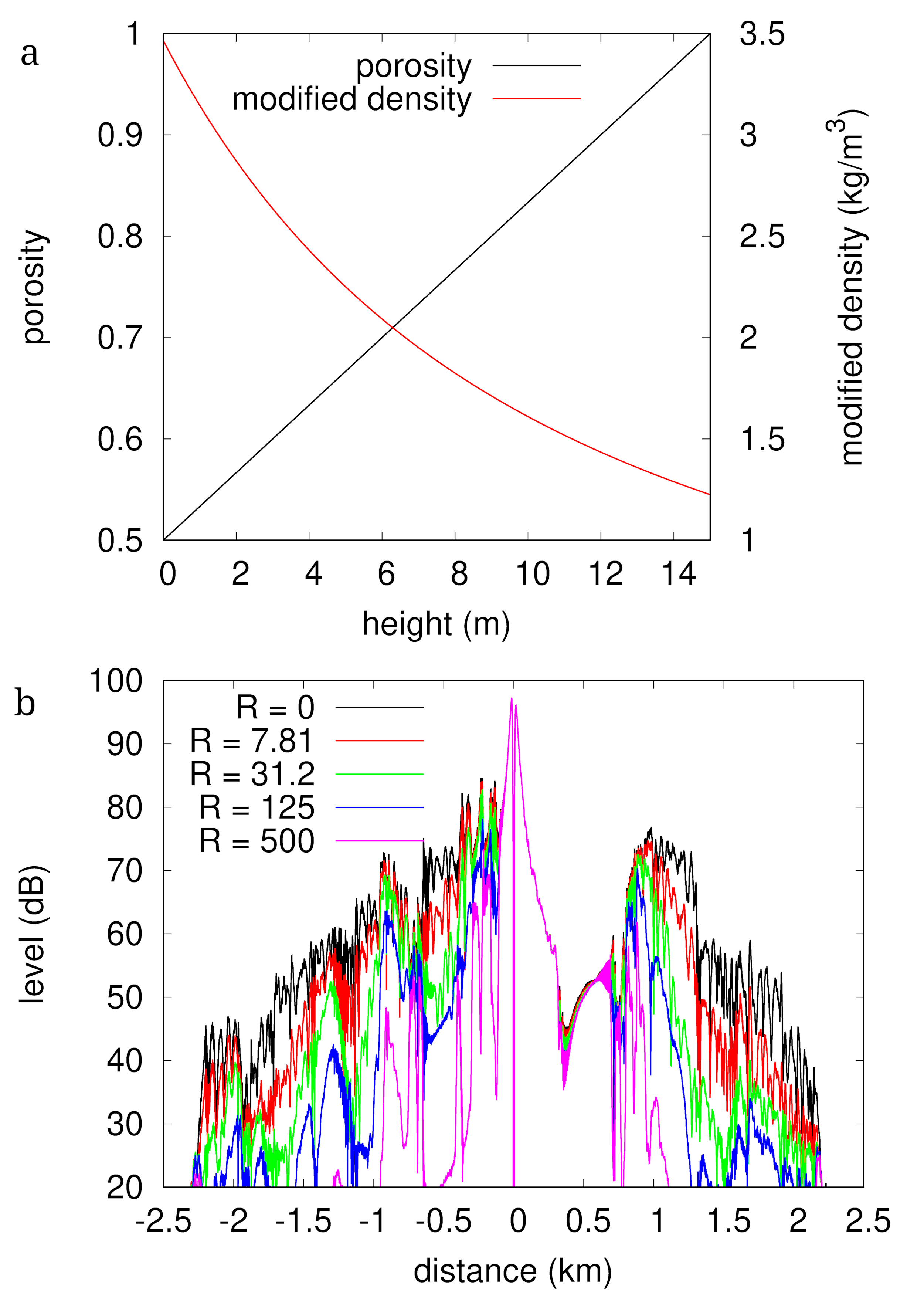} 
	\caption{The effects of vegetation on sound levels. a) The porosity and modified density as a function of height that is assumed in areas where trees are present. b) The effects of flow resistivity, which varies linearly from these values at the floor to zero at \qty{15}{\m}, on the sound level as a function of distance from the site.}
\end{center}
\end{figure}

The effects of vegetation is now considered. Figure 3 depicts the porosity and modified density in the current model as a function of height.
In the current model, we assume the porosity increases linearly from 0.5 to 1 as the height goes from 0 to \qty{15}{\m}.
Recall, that the tortuosity is directly calculated from the porosity. 
The modified density, which is proportional to tortuosity and inversely proportional to porosity, increases significantly as the height is reduced.
The effects of including vegetation in the current model is depicted in Fig. 2b. 
The flow resistivity is varied from \qty{0}{\kilo\Pa\s\per\square\m} to \qty{500}{\kilo\Pa\s\per\square\m}, with the attenuation of sound significantly dependent on this parameter. 
It is worth noting, however, that because the \qty{100}{\Hz} sound waves have a wavelength of around \qty{3.4}{\m}, the sound waves may pass through the vegetative canopy relatively unhindered as the wavelength is larger than the size of individual leaves and twigs.
Therefore, we might expect the value of the flow resistivity in this system to be small.

\begin{figure}[h]
\begin{center}
	\includegraphics[width=0.55\linewidth, alt = {Two plots. The top plot shows the sound level as a function of distance. The second plot is the exact same data, except zoomed in to show a close up of the difference as temperature variations only account for a change of several decibels.}]{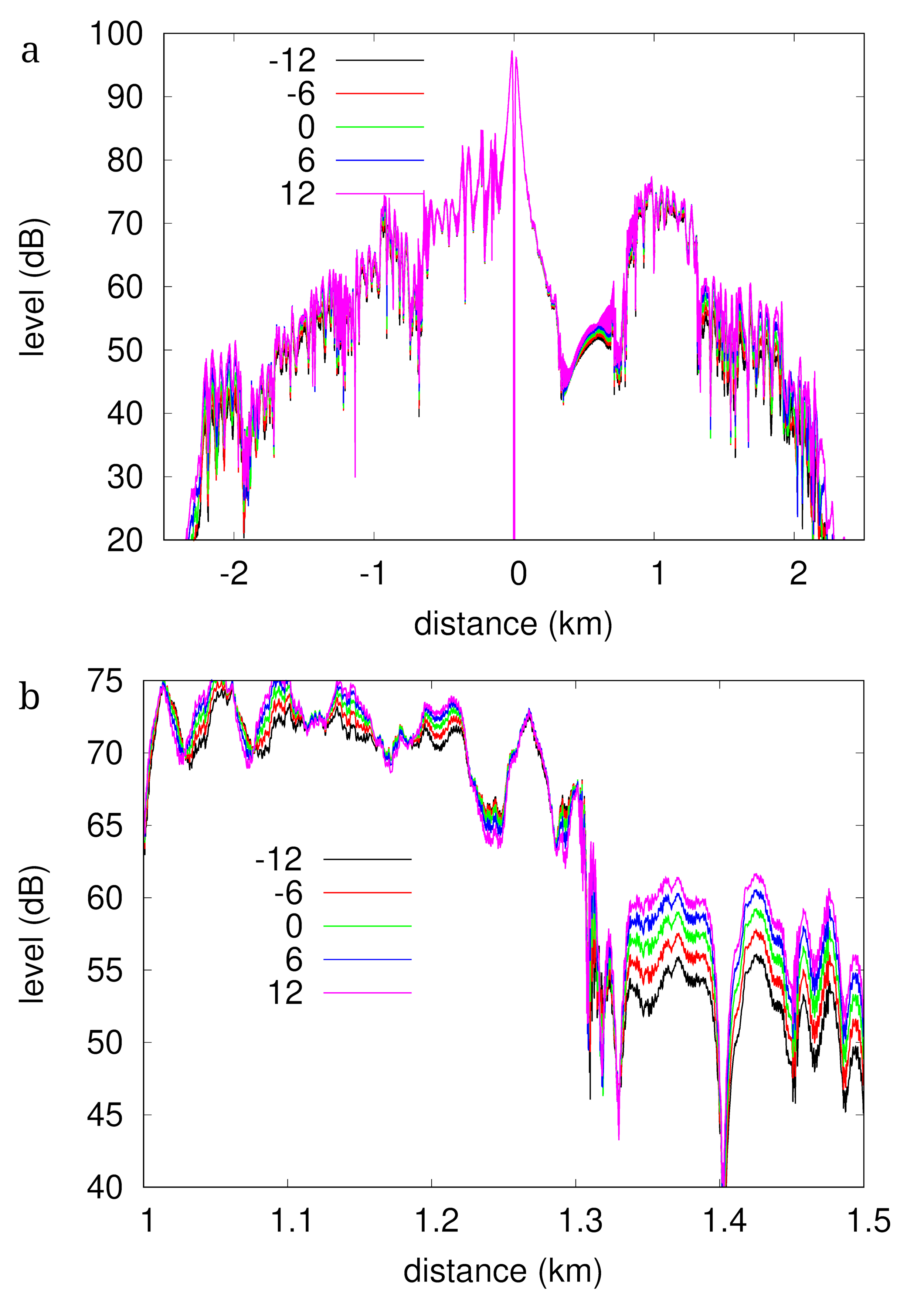} 
	\caption{The effects of temperature gradient in the air. a) The sound level as a function of distance from the site as a function of temperature gradients from \qty{-12}{\degreeCelsius\per\km} to \qty{12}{\degreeCelsius\per\km}. b) The sound level as a function of distance from the site showing the small changes in dB predicted as a function of temperature gradients at a location across the river from the AI data center.}
\end{center}
\end{figure}

The effects of temperature gradients in the atmosphere appear to be less pronounced. During the daytime the ground is heated by the sun, making the air near the surface warmer than higher up (negative gradient). While at night the ground cools via the emission of infrared radiation, making the air above it warmer (positive gradient). 
Figure 4 shows the effects of including temperature gradients from \qty{-12}{\degreeCelsius\per\km} to \qty{+12}{\degreeCelsius\per\km}, which represents a relatively large range of gradients \cite{bolstad1998measured}.
Fig. 4a reveals a relatively small predicted impact of these temperature gradients on the noise pollution. 
However, a closer look (in Fig. 4b) reveals that local variations of several decibels might occur as a consequence of temperature gradients. 
The positive gradient, which causes acoustic refraction to bend sound waves back towards the surface rather than allowing them to propagate upwards and away from the surface, locally increases noise pollution by around 2--\qty{3}{dB}.

\section*{Conclusions}

The propagation of a low frequency sound waves from the site of an AI data center in Beaver County has been simulated using the FDTD method. 
The numerical predictions indicate that in regions close to the AI data center, and regions of sudden elevations on either side of the river, the noise pollution may be significant.
Mitigation strategies may be required to reduce the impact of noise pollution to local residents.
However, because of the unique topography, the noise pollution may be insignificant beyond these regions.
For example, regions of Raccoon Township, PA, that are over the intervening escarpment are predicted to experience relatively small noise pollution.
The effects of vegetation are difficult to estimate, but are not expected to be significant, while temperature gradients may locally increase noise pollution by a couple of decibels.

\bibliographystyle{unsrt} 
\bibliography{paper.bib}

\end{document}